\documentclass[preprint,superscriptaddress,preprintnumbers,showpacs]{revtex4}
\usepackage{amssymb}
\usepackage{graphicx}
\usepackage{dcolumn}
\usepackage{bm}

\begin{document}

\newcommand*{\sjtu}{INPAC, Department of Physics and Shanghai Key Laboratory for Particle Physics and Cosmology, Shanghai Jiao Tong University,  Shanghai, China}\affiliation{\sjtu}
\newcommand*{\ncts}{Department of Physics, National Tsing Hua University, and National Center for Theoretical Sciences, Hsinchu, Taiwan}\affiliation{\ncts}
\newcommand*{\NTU}{CTS, CAST and Department of Physics, \\National Taiwan University, Taipei, Taiwan}\affiliation{\NTU}

\hfill{CETUP*-12/005}

\title{Radiative Two Loop Inverse Seesaw
and Dark Matter}

\author{Gang Guo}\affiliation{\sjtu}
\author{Xiao-Gang He}\email{hexg@phys.ntu.edu.tw}\affiliation{\sjtu}\affiliation{\ncts}\affiliation{\NTU}
\author{Guan-Nan Li}\affiliation{\sjtu}

\begin{abstract}
Seesaw mechanism provides a natural explanation of light neutrino masses through suppression of heavy seesaw scale.
In inverse seesaw models the seesaw scale can be much lower than that in the usual seesaw models. If terms inducing
seesaw masses are further induced by loop corrections, the seesaw scale can be lowered to be in the range probed by
experiments at the LHC without fine tuning. In this paper we construct models in which inverse seesaw neutrino masses are generated at
two loop level. These models also naturally have dark matter candidates. Although the recent data from Xenon100 put stringent constraint on the
 models, they can be consistent with data on neutrino masses, mixing,
dark matter relic density and direct detection. These models also have some interesting experimental
signatures for collider and flavor physics.

\end{abstract}


\date{\today $\vphantom{\bigg|_{\bigg|}^|}$}

\maketitle

\section{Introduction}

Seesaw mechanism is one of the popular mechanisms\cite{seesawI,seesawII,seesawIII} beyond the standard model (SM)
which can provide some explanations why
neutrino masses are so much smaller than their charged lepton partner mass scales $m_D$.
In Type I and III seesaw models\cite{seesawI,seesawIII} it requires the existence of heavy right-handed neutrinos
of a Majorana mass scale $M$. The light neutrino mass is of order $m_D (m_D/M)$. The usual
scale of the heavy right-handed neutrino mass $M$ is expected to be super heavy which can
be as high as the grand unification scale.  It would be good if the seesaw mechanism can
be tested by high energy colliders. The LHC can test theoretical models
beyond the SM at an energy scale as high as 8 TeV at present and will reach 14 TeV in the
future. If indeed the heavy seesaw scale is of grand unification scale, it is impossible to test
seesaw mechanism directly at accessible collider energies. Theoretically it is interesting
to see if the seesaw scale can be lowered to TeV range allowing direct probe of ATLAS and CMS
experiments at the LHC. There are indeed special solutions which allow lower heavy right
handed neutrinos of order TeV with the price of fine tuning of the parameters\cite{low-seesaw}. Although
this is theoretically allowed, it loses the original motivation of naturally explanation
for the lightness of neutrinos through seesaw mechanism. Radiative seesaw neutrino mass
generation can easy the problem and at the same time provide the much desired candidate
for dark matter when additional symmetry exists to stablize the dark matter candidate\cite{ma,he-ren}.
The inverse seesaw mechanism\cite{inverse-seesaw} can also lower the seesaw scale. In the inverse seesaw model,
the heavy Majorana neutrinos are replaced by the heavy Dirac particles. The light neutrino
masses are of order $\mu (m_D/M)^2$. Here $M$ is the heavy Dirac particle scale and $\mu$
is a Majorana mass of the heavy Dirac particles which are supposed to be small even compared
with $m_D$. It is clear that the heavy Dirac mass scale $M$ can be much lower than that for the
Majorna mass in the usual seesaw models naturally. If the inverse seesaw is also achieved by
radiative correction, the heavy scale can be even lower\cite{ma1,sandy,bazzocchi}. There are also other mechanisms to
further lower the scale by naturally having a small $\mu$ parameter, such as that discussed in Ref.\cite{fong} through extra
warped dimension. Here we will study radiative inverse seesaw models.
In achieving radiative mass generation,
sometimes it involves introduction of new symmetries to forbid terms which may induce tree level
neutrino masses. If the symmetry introduced is unbroken, there may be a stable new particle in the
theory. This new particle may play the role of the dark matter needed to explain about 23\%
of the energy budget of our universe\cite{dark-matter}.
In this paper we study several simple
models using a leptonic heavy Dirac multiple to facilitate radiative inverse seesaw neutrino
mass and also to have dark matter candidate.
\\

\section{ Tree Inverse Seesaw}

The inverse seesaw neutrino mass matrix $M_\nu$ is the mass matrix resulted from the effective Lagrangian
\begin{eqnarray}
L_m = - \bar \nu_L m_D N_R  -  \bar N_L M N_R -  {1\over 2} \bar N^c_R \mu_R N_R  - {1\over 2} \bar N_L \mu_L N^c_L  + h.c.
\end{eqnarray}
where $\nu_L$ is the light active neutrino, $N_{L,R}$ are heavy neutrinos.

In the bases $(\nu^c_L, N_R, N_L^c)^T$, $M_\nu$ is given by
\begin{eqnarray}
M_\nu = \left ( \begin{array}{ccc}
0&m_D&0\\
m_D^T&\mu_R&M^T\\
0&M&\mu_L
\end{array} \right )\,
\end{eqnarray}

With the hierarchy $\mu_L\sim \mu_R << m_D << M$, the light neutrino mass matrix $m_\nu$, defined by
$L_{mass} = - (1/2)\nu_L m_\nu \nu^c_L$, to order $(m_D/M)^2$ is given by\cite{inverse-seesaw}
\begin{eqnarray}
m_\nu = m_DM^{-1}\mu_L (M^{-1})^Tm_D^T.
\end{eqnarray}

There are different ways to achieve inverse seesaw mechanism depending on where $N_{L,R}$ comes from.
We briefly outline two simple possibilities which may realize inverse seesaw at tree level.

One of the simplest ways is to introduce right-handed $N_R$ and left-handed $N_L$ singlet
heavy neutrinos with a discrete $N_R \to  N_R$, $N_L \to  - N_L$ $Z_2$ symmetry and all
other SM particles do not transform under this symmetry. The $m_D$ term is generated through Yukawa
coupling $\bar L_L Y_D \tilde H N_R$. Here $H = (h^+, (v_H+ h+ iI)^T/\sqrt{2}$ is the SM Higgs doublet
which transform under the SM electroweak gauge group $SU(2)_L\times U(1)_Y$ as $(2,1/2)$. $\tilde H = i\sigma_2 H^*$.
$v_H$ is the vacuum expectation value (vev) of $H$. $L_L = (\nu_L, e_L)^T: (2,-1/2)$ is the SM lepton doublet.
The $\mu_{L,R}$ terms are from bare Majorana mass terms $\bar N^c_R \mu_R N_R$ and  $\bar N_L \mu_L N^c_L$.
Because the $Z_2$ symmetry, the bare Dirac mass term $\bar N_L M N_R$ is not allowed. In order to generate
a non-zero M term, one can introduce a singlet scale $S$ transforming
under $Z_2$ as $S\to - S$ with a non-zero vev $v_s/\sqrt{2}$. In this case, the Yukawa term $\bar N_L Y_s S N_R$
is allowed which generates a $M$ given by $Y_s v_s/\sqrt{2}$.

One can also introduce a leptonic doublet $D_{L,R}: (2,-1/2)$ along with a singlet $S$ and a triplet
$\Delta: (3, - 1)$ ($\Delta_{ij}$ with $\Delta_{11} = \Delta^0$, $\Delta_{12}=\Delta_{21} =
\Delta^{-}/\sqrt{2}$ and $\Delta_{22} = \Delta^{--}$) to realize the inverse seesaw.
One can introduce a global $U(1)_D$ symmetry to distinguish $D_L$ and $L_L$. Under this symmetry
$D_{L,R} \to exp[i\alpha_D]D_{L,R}$, $S \to exp[-i\alpha_D]S$,
$\Delta \to exp[2i\alpha_D] \Delta$, and other fields do not transform. We have the following
Lagrangian relevant to neutrino masses
\begin{eqnarray}
L_D = - \bar L_L Y_D D_R S - \bar D_L M D_R - {1\over 2} \bar D_L Y_L D^c_L \Delta - {1\over 2} \bar D_R^c Y_R D_R \Delta^\dagger + h.c.
\end{eqnarray}

If both $S$ and $\Delta$ develop non-zero vev's, the inverse seesaw mechanism is realized. This model, however,
will have a Goldstone boson due to breaking of the global $U(1)_D$ symmetry which may be problematic.
To avoid the existence of a Goldstone boson in the theory, extension is needed. Also in the above two models, no candidates for dark matter.

In following sections, we will extend the two models discussed in this section to radiatively generate
inverse seesaw neutrino masses. We will discuss the
possibility of having dark matter candidates in these models.
\\

\section{ Radiative Two Loop Inverse Seesaw}

To avoid the appearance of massless Goldstone boson in the theory, a possible approach is not to allow the global
symmetry to break and therefore no Goldstone boson emerges.
Applying this idea to the model involving $D_{L,R}$, $S$ and $\Delta$ are then not allowed to have vev's. This however also firbids the light neutrinos
to have non-zero masses at tree level. We have to extend the model.
To this end, we introduce another singlet $\sigma$ which transforms under the $U(1)_D$ as $\sigma \to exp[2i\alpha_D]\sigma$.
We refer to this model as the $U(1)_D$ model.
The allowed renormalizable terms in the potential $V_D$ are given by
\begin{eqnarray}
V_D &=& -\mu^2_H H^\dagger H + \lambda_H (H^\dagger H)^2 + \mu^2_S S^\dagger S + \lambda_S (S^\dagger S)^2 + \mu^2_\sigma \sigma^\dagger \sigma + \lambda_\sigma (\sigma^\dagger \sigma)^2\nonumber\\
& + & \mu^2_\Delta \Delta^\dagger \Delta + \lambda^\alpha_\Delta (\Delta^\dagger \Delta \Delta^\dagger \Delta)_\alpha + \sum_{ij}\lambda_{ij} i^\dagger ij^\dagger j
+ (\mu_{S\sigma} S^2 \sigma  + \lambda_{\Delta \sigma H} H \Delta \sigma^\dagger H + h.c.),
\end{eqnarray}
where the sum $\sum_{ij}$ is over all possible $i$ and $j$, and $i$ to be one of the $H$, $S$, $\sigma$ and $\Delta$.
The allowed terms are:
\begin{eqnarray}
\lambda^\beta_{H\Delta} (H^\dag H \Delta^\dag \Delta)_\beta + \lambda_{H\sigma} (H^\dag H \sigma^\dag \sigma)+ \lambda_{HS} (H^\dag H S^\dag S)
+ \lambda_{\Delta \sigma} (\Delta^\dag \Delta \sigma^\dag \sigma) + \lambda_{\sigma S} (\sigma^\dag \sigma S^\dag S)\;.
\end{eqnarray}
In the above the indices $\alpha$ and $\beta$ indicate different ways of forming singlet. They are given by
\begin{eqnarray}
  (\Delta^\dag \Delta \Delta^\dag \Delta)_1 = \Delta^*_{ij} \Delta_{ij} \Delta^*_{kl} \Delta_{kl}\;,\;\;\;\;(\Delta^\dag \Delta \Delta^\dag \Delta)_2 = \Delta^*_{ij} \Delta_{ik} \Delta^*_{kl} \Delta_{jl} \cr
  (\Delta^\dag \Delta H^\dag H)_1 = \Delta^*_{ij} \Delta_{ij} H^*_{k} H_{k}\;,\;\;\;\;(\Delta^\dag \Delta H^\dag H)_2 =  \Delta^*_{ij} \Delta_{kj} H^*_{k} H_{i}
\end{eqnarray}

In the above $\mu^2_i$ are all larger than zero. The potential only allows $H$ to have a non-zero vev $v_H$. The theory has an unbroken $U(1)_D$ global symmetry after spontaneous
symmetry breaking from $SU(2)_L\times U(1)_Y$ to $U(1)_{em}$. At the tree level, light neutrinos
are massless. Giving the above terms in $L_D$ and $V_D$, it is not possible to defined conserved lepton number. This is because that among the $\bar L_L D_R S$, $\bar D_{L,R} D^c_{L,R} \Delta$ and $S^2\sigma$, and $H \Delta \sigma^\dagger H$ vertices, there is always one vertex where lepton number is violated. For example, assigning $L_L$ to have lepton number $+1$ and $D_{L,R}$ to have $X$, if one demands conservation of lepton number as $\bar L_L D_R S$, $\bar D_{L,R} D^c_{L,R} \Delta$ and $S^2\sigma$ vertices, at the vertex $H \Delta \sigma^\dagger H$ would violate lepton number by 2 units. One can demand other vertices to converse lepton number, but no matter what one chooses, the combination of terms proportional to $Y_DY_{L} \mu_{S\sigma}\lambda_{\Delta \sigma H}Y_D$ always violate lepton number by 2 units. However, the global $U(1)_D$ symmetry is respected. Because of lepton number is violated, at loop levels, Majorana neutrino masses may be generated.
We find that non-zero Majorana neutrino masses can be generated at two loop level shown in Fig.1. This two loop contribution violates the lepton number by 2 units. This two loop mass generation is similar to the Babu-Zee model two loop neutrino mass generation\cite{babu-zee} but with the light charged leptons in the loop replaced by new heavy particles.
The last two terms in the potential are crucial for light neutrino mass generation.

\begin{figure}
\includegraphics[width=6cm]{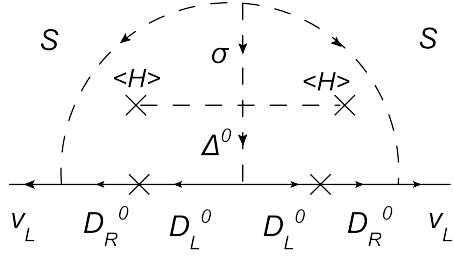}
\caption{Two loop diagram for neutrino mass generation.}\label{two-loop}
\end{figure}

We will now describe how to calculate the two-loop induced light neutrino mass. After $H$ develops vev,
mixing between $\Delta^0$ and $\sigma$ will be generated via the term $ H \Delta \sigma^\dagger H$. The corresponding mass matrix for $(\Delta^0, \sigma)^T$ can be expressed as:
       \begin{eqnarray}
       \left(
       \begin{array}{cc}
       M^2_{11} & M^2_{12} \\
       M^2_{21} & M^2_{22}
       \end{array}
       \right)          \nonumber
       \end{eqnarray}
  with
        \begin{eqnarray}
        M^2_{11} &=& \mu^2_\Delta + {1\over 2}\lambda^1_{H\Delta} v^2_H\;,\;\; M^2_{22} = \mu^2_\sigma + {1\over 2}\lambda_{H\sigma} v^2_H\;,\;\;M^2_{12} =  M^2_{21} = {1\over 2} \lambda_{\Delta \sigma H } v^2_H\;.
        \end{eqnarray}

One can diagonalize the mass matrix via:
   \begin{equation}
       \left(\begin{array}{cc}
       \phi_1    \\
       \phi_2
       \end{array}
      \right) =
       \left(
       \begin{array}{cc}
       \cos \alpha & \sin \alpha \\
       -\sin \alpha & \cos \alpha
       \end{array}
       \right)
      \left(
       \begin{array}{cc}
       \Delta^{0}    \\
       \sigma
       \end{array}
       \right)              \nonumber
      \end{equation}
    with
       \begin{eqnarray}
       m^2_{\phi_{1,2}} &=& {  M^2_{11} + M^2_{22} \pm \sqrt{ (M^2_{11} - M^2_{22} )^2 + 4 M^2_{12} M^2_{21} }   \over 2 }\;,   \cr
       \sin 2\alpha  &=&  {  2 M^2_{12}  \over  \sqrt{ (M^2_{11} - M^2_{22} )^2 + 4 M^2_{12} M^2_{21} } }\;.  \nonumber
       \end{eqnarray}

    The light neutrino mass $m_\nu$ generated via two-loop diagram is given by:
  \begin{eqnarray}
  m_\nu &=& Y_D^2 M^2  Y_L  \mu_{S\sigma} \cos\alpha \sin\alpha  \int {d^4p\over (2\pi)^4} {d^4q \over (2\pi)^4}{ 1 \over p^2-m_S^2 } { 1 \over q^2-m_S^2 }{ 1 \over p^2-M^2 }{ 1 \over q^2-M^2 }\nonumber\\
   &\times& \left ({ 1 \over (p-q)^2-m_{\phi_1}^2 } - { 1 \over (p-q)^2-m_{\phi_2}^2 }\right )\;.
  \end{eqnarray}

   The last factor in the above can be written as $(m^2_{\phi_1} - m^2_{\phi_2})/((p-q)^2-m^2_{\phi_1})((p-q)^2-m^2_{\phi_2})$. Using,  $\sin\alpha \cos\alpha (m^{2}_{\phi_1}-m^{2}_{\phi_2}) = M^2_{12}={1\over 2}\lambda_{\Delta\sigma H}v_H^2$, and neglecting the mass splitting between $m_{\phi_1}$ and $m_{\phi_2}$ in the denominator, we obtain
   \begin{eqnarray}
   m_\nu=&&\frac{\lambda_{\Delta\sigma H} Y_L Y_D^2 \mu_{S\sigma} v_H^2 M^2}{2(M^2-m^2_{S})^2}\int\frac{d^{4}p}{(2\pi)^4}\frac{d^{4}q}{(2\pi)^4}\times\nonumber\\
   &&\frac{1}{[(p-q)^2-m^2_{\phi_1}]^2}(\frac{1}{p^2-m^2_{S}}\frac{1}{q^2-m^2_{S}}-\frac{1}{p^2-m^2_{S}}\frac{1}{q^2-M^2}-\frac{1}{p^2-M^2}
   \frac{1}{q^2-m^2_{S}}\nonumber\\
   &&+\frac{1}{P^2-M^2}\frac{1}{q^2-M^2})\;.
  \end{eqnarray}
Our results for the two loop integral agree with that obtained in Ref.\cite{bruce}.

Carrying out the loop integrals, we finally obtain
   \begin{eqnarray}
   m_\nu=&&\frac{\lambda_{\Delta\sigma H} Y_L Y_D^2 \mu_{S\sigma} v_H^2 }{2(4\pi)^4 M^2(1-m^2_S/M^2)^2} [g(m_{\phi_1},m_S,m_S)-g(m_{\phi_1},M,m_S)\cr
   &&-g(m_{\phi_1},m_S,M)+g(m_{\phi_1},M,M)]\;,
   \end{eqnarray}
where
  \begin{eqnarray}
  g(m_1,m_2,m_3)=\int^{1}_{0} dx [1+Sp(1-\mu^2)-\frac{\mu^2}{1-\mu^2}\log\mu^2]\nonumber
  \end{eqnarray}
  with $\mu^2=\frac{ax+b(1-x)}{x(1-x)}, a=\frac{m^2_2}{m^2_1}, b=\frac{m^2_3}{m^2_1}$.
$Sp(z)$ is the Spence function or the dilogarithm function defined as:
  \begin{eqnarray}
  Sp(z)=-\int^z_0 {\ln(1-t)\over t} dt
  \end{eqnarray}

To compare with the inverse seesaw mass formula, we rewrite the above in a matrix form in the bases where $M$ is diagonalized
  \begin{eqnarray}
  m_\nu^{ij} =\frac{v_H Y_D^{ik}(\lambda_{\Delta\sigma H}\mu_{S\sigma} Y_L^{kl}) Y_D^{jl}  v_H}{M^2_{kk}}\kappa_{kl}\;, \label{mass}
  \end{eqnarray}
 where $\kappa_{kl}$ is defined as:
 \begin{eqnarray}
 \kappa_{kl} &=&\delta_{kl}\frac{1}{2(4\pi)^4}{1\over (1- m^2_S/M^2_{kk})^2}[g(m_{\phi_1},m_S,m_S)-g(m_{\phi_1},M_{kk},m_S) \cr
 &&-g(m_{\phi_1},m_S,M_{kk})+g(m_{\phi_1},M_{kk},M_{kk})]
 \end{eqnarray}

If one identifies, effectively, $m_D = Y_D v_H$, $M = diag(M_{ii})$ and $\mu_L = (\mu_L^{ij})$ with $\mu_L^{ij} = (\lambda_{\Delta\sigma H}\mu_{s\sigma}) Y_L^{ij}\kappa_{ij}$,
the light neutrino mass matrix is effectively an inverse seesaw mass form. We therefore refer this as radiative inverse seesaw mechanism. This model is different than those
radiative inverse seesaw models discussed in Ref.\cite{ma1,sandy} where additional neutral heavy spin-half particles are introduced to generate radiative neutrino masses.

The above formula can easily fit current data on neutrino mixing and masses\cite{neutrino-data}. As an example, let us consider a simple case with $Y_L$ diagonal and $Y_D =y_D U_{PMNS}$. For the normal hierarchy, choose $Y_L = diag(1, ~1.05, ~
2.01)\times 10^{-2}$, $y_D = 10^{-2}$, $\lambda_{\Delta\sigma H} =0.1$, $\mu_{S\sigma} = 100$GeV, $m_{\phi_1} = 300$GeV, $m_S = 150$GeV, $M_{ii}=500$GeV, we can get all the three neutrino mass $2.804\times 10^{-2}$eV, $2.936\times 10^{-2}$eV, $5.636\times 10^{-2}$eV, respectively. These are consistent with data.
For inverted hierarchy case, we just need to replace $Y_L$ with  $Y_L = diag(1.297, ~1.317, ~0.100)\times 10^{-2}$, with all the other parameters unchanged, the neutrino masses will be $4.90\times 10^{-2}$eV, $4.98\times 10^{-2}$eV, $3.78\times 10^{-3}$eV, respectively. Again, these numbers are consistent with data.

Along the same idea, the case with singlet heavy neutrinos discussed earlier, can also be modified to have two loop realization of inverse seesaw
mechanism. To this end we impose on the theory a global $U(1)_S$ symmetry. The new particles beyond SM are: $N_{L,R}: (1,0)$, $\eta: (2,-1/2)$, $\Delta: (3, -1)$ and $S:(1,0)$. Under the $U(1)_S$
these particles transform as: $N_{L,R} \to exp[i\alpha_S]$, $\eta \to exp[-i\alpha_S] \eta$, $\Delta \to exp[-2i\alpha_S]$, $S \to exp[-2i\alpha] S$.
The Lagrangian $L_S$ for the bare mass term and Yukawa couplings, and the potential $V_S$ relevant for two loop neutrino mass generation are given by
\begin{eqnarray}
L_S &=& - \bar N_L M N_R - \bar L_L Y_D N_R \eta - {1\over 2} \bar N^c_R Y_R N_R S - {1\over 2} \bar N_L Y_L N^c_L S^\dagger + h.c.\nonumber\\
V_S&=& \mu_{\Delta \eta}\eta \Delta^\dagger \eta  + \lambda_{\Delta S H} H \Delta^\dagger S H + h.c. + ...,
\end{eqnarray}
where ``...'' indicate other allowed terms.

The light neutrino mass matrix can be obtained by replacing $\mu_{S\sigma}$ by $\mu_{\Delta\eta}$, and $\lambda_{\Delta \sigma H}$ by $\lambda_{\Delta S H}$ in eq.\ref{mass}. In this
model terms proportional to $Y_D Y_L\mu_{\Delta\eta}\lambda_{\Delta SH} Y_D$ violates lepton number by 2 units, for the same reasons for the $U(1)_D$ model.

As long as the radiative generation of inverse seesaw neutrino masses is concerned the above two models are very similar. However, when considering dark matter physics,
these two models have different features. We proceed to discuss them in the following.
\\

\section{ Dark Matter Candidate}

Since in both the $U(1)_D$ and $U(1)_S$ models, the global symmetries are not broken, there are stable particles which may play the role of dark matter. Which one of the new particles is the lightest one depends on the parameter space and therefore determines which one plays the role of dark matter.

In the $U(1)_D$ model, the heavy fermion particles have non-zero hypercharge and cannot play the role of dark matter. This is because that although dark matter relic density can be produced by dark matter annihilate into gauge particle with known interaction strength with sufficiently large dark matter mass, the direct detection rate from t-channel Z boson exchange would be too large. This possibility is therefore ruled out.
The neutral components of the scalar fields in the models are other possibilities which may be identified as dark matter. The neutral component $\Delta^0$,
has problem to play the role of dark matter due to its non-zero hypercharge. If the real and imaginary parts of the $\Delta^0$ masses $m_r$ and $m_i$ have a splitting
$\delta = m_r - m_i$, the non-zero hypercharge problem can be resolved by invoking the inelastic dark matter mechanism\cite{inelastic-dm}, namely the scattering of a dark matter
off nucleon is kinematically forbidden if the mass splitting $\delta$ is larger than 100 KeV or so.
In the $U(1)_D$ model, however, we find that it is not possible to generate a non-zero $\delta$ for the real and imaginary parts in $\Delta^0$, the inelastic dark matter mechanism is ineffective.

The natural dark matter field is $S$. It does not have a non-zero hypercharge
and does not mix with any particle having hypercharge. As long as dark matter properties are concerned, this model is very similar to the real
singlet (darkon) model\cite{real-singlet} and therefore similar dark matter properties\cite{darkon} and identical to the complex scalar singlet model\cite{complex-singlet} with degenerate mass for the real and imaginary parts of S.
This is a typical Higgs portal model. Dark matter annihilation and detection are all mediated by Higgs boson.

The term important is $S^\dagger S H^\dagger H$. Removing the would-be Goldstone bosons in $H$, we have
\begin{eqnarray}
\lambda_{SH} S^\dagger S H^\dagger H = {1\over 2}\lambda_{SH}(v^2_H + 2v_H h + hh)SS^\dagger\;.
\end{eqnarray}

The first term will modify the mass of $S$ from $\mu^2_S$ to $M_D^2 = \mu^2_S + \lambda_{SH} v^2_H/2$. As long as dark matter annihilation and detection are concerned,
the free parameters are: $M_D$, $\lambda_{SH}$ and also the Higgs boson mass. In the model, the Higgs boson $h$ properties, its mass and its couplings to SM particles (fermions and gauge bosons), are very close to the SM Higgs boson $h_{SM}$. The recent LHC data indicate that the mass is about 125 GeV\cite{LHC-higgs}. We will analyze the model using Higgs mass of $m_h = 125$ GeV. The dark matter relic density and direct detection constraints on the coupling $\lambda_{SH}$ and $M_D$ are shown in Fig. 2. Since now we have two degenerate components as dark matter, the constraint on $\lambda_{SH}$ from relic density is $1/\sqrt{2}$ times smaller than the darkon model\cite{real-singlet}. The recent data on direct dark matter search from Xenon100\cite{newxenon100} put the most strigent constraint on the allowed range for dark matter mass. The range of a few tens of GeV for dark matter mass is in trouble. However, dark matter mass about half of the Higgs mass and larger than 130 GeV is still allowed.

\begin{figure}
\includegraphics[width=6cm]{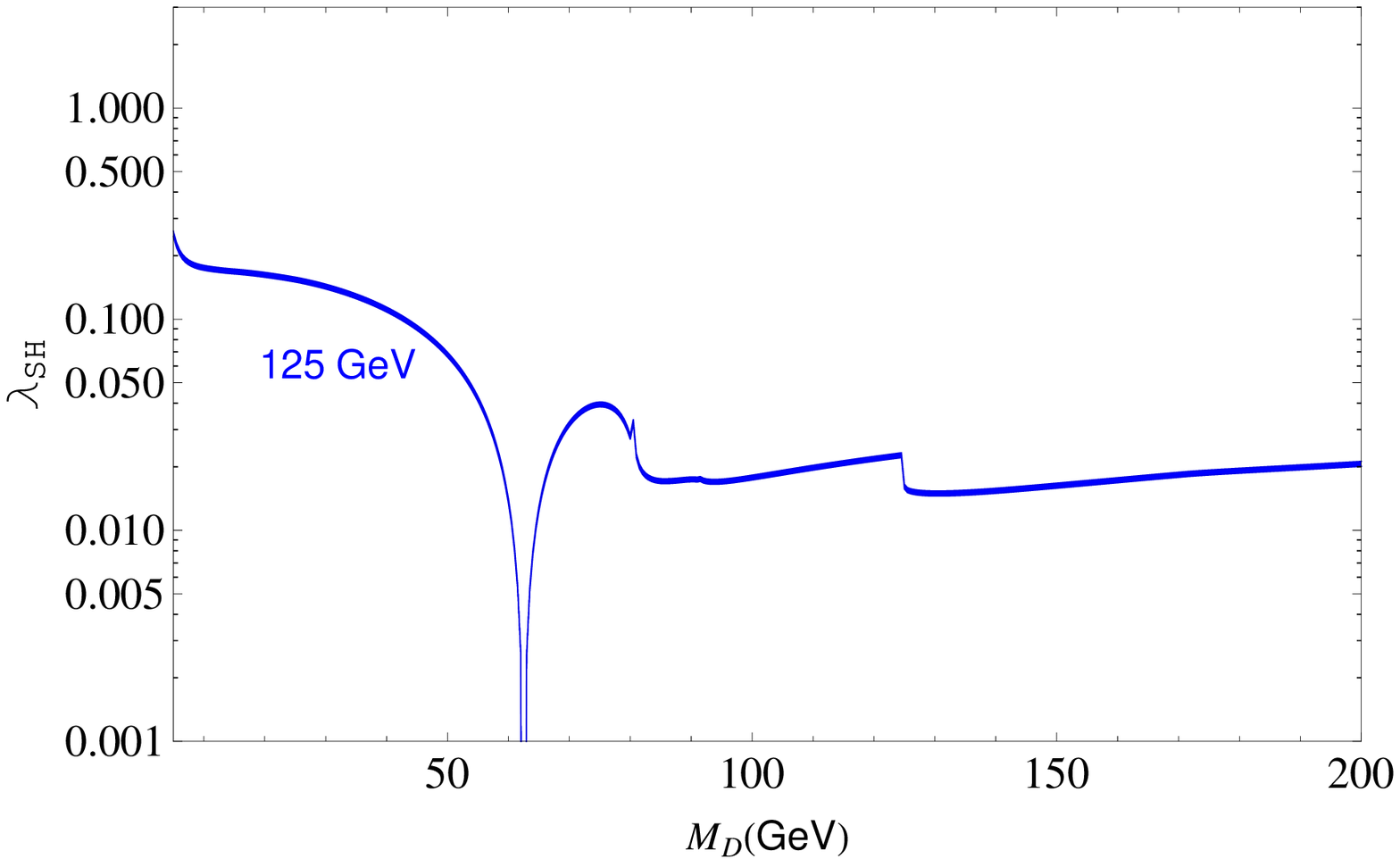}\\
\includegraphics[width=6cm]{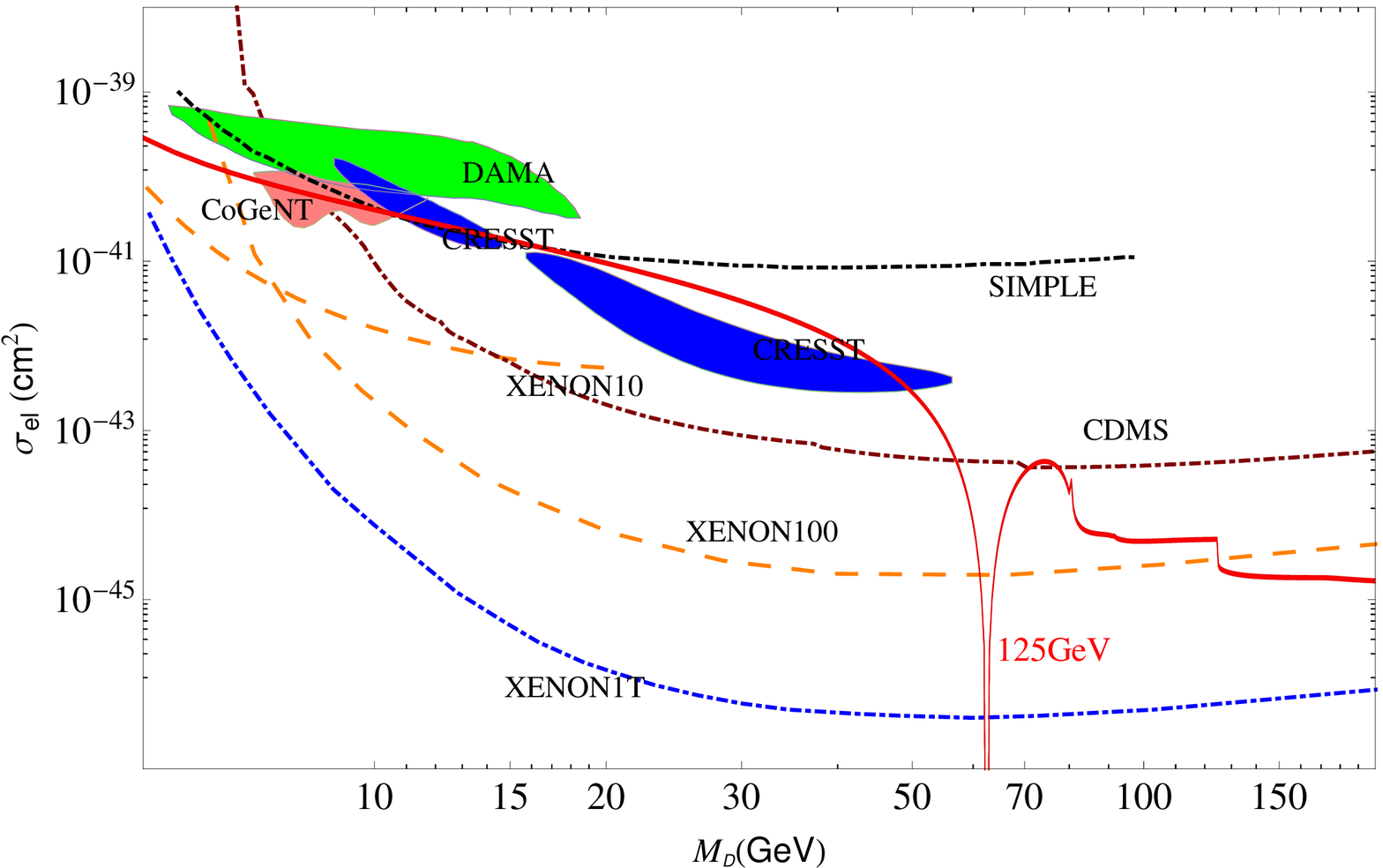}
\caption{Constraints on the coupling and dark matter mass from dark matter relic density and direct detection\cite{dark-direct,newxenon100} for $S$ as the dark matter with Higgs mass set to be 125 GeV. The projected Xenon1T sensitivity is also drawn.}\label{S-dark-matter}
\end{figure}

The $\sigma$ field is also a possibility for dark matter since it does not have a hypercharge neither. It  mixes with $\Delta^0$ after $H$ develops $vev$ through the
term: $\lambda_{\Delta \sigma H} H \Delta \sigma^\dagger H$. The lighter of physical particle which may play the role of dark matter will also has a
non-zero $Z$ coupling. However, in this case there is the mixing parameter to tune to satisfy the constraint. We find that as long as the parameter $\sin \alpha$ is less than $10^{-3}$, the large direct detection cross section can be solved. We also checked that $\sin\alpha$ of order $10^{-3}$ can be made compatible with the neutrino mass generation requirement. With $\alpha < 10^{-3}$, the dark matter is dominated by the component $\sigma$. The dark matter properties are similar to $S$.

We now briefly discuss dark matter properties in the $U(1)_S$ model. In this model, the neutral scalars in $\eta$, $\Delta$ and $S$, and the $N$ are possible candidates for dark matter. The
neutral components in $\eta$ and $\Delta$ have hypercharges and also there are no mass splitting between the real and imaginary parts, they
have too large cross section for direct dark matter detection after fitting relic density requirement. The $S$ field although does not have a hypercharge, it mixes with $\Delta^0$, some fine tuning is needed to be compatible
with direct dark matter detection data. The situation is similar to the case of $\sigma$ as the dark matter in the $U(1)_D$ model. This is similar to the case of $\sigma$ as dark matter as in the $U(1)_D$ model.

The $N$ field does not have hypercharge and may also play the role of dark matter. In this case, the dark matter relic density is achieved by t-channel exchange of $\eta$ induced $N\;N$ pair annihilate into lepton pairs, $l^+\;l^-$ and $\nu_L\; \bar\nu_L$. The annihilation rate is governed by the Yukawa coupling $Y_D$, the mass $m_\eta$ of $\eta$ and also the dark matter mass $M_D = M$.
We have checked that there are parameter space where the correct relic density can be produced. In Fig.3 we show some correlations of the parameters which can produce the correct relic density. At the tree level, $N$ does not couple to quarks. However, at one loop level, with $L_L$ and $\eta$ in the loop $\bar N$-$N$-$Z$ coupling can be generated which can lead to sizeable dark matter direct detection cross section and at the same time satisfy the dark matter relic density constraint. The results are shown in Fig. 3. We again see that the recent Xenon100\cite{newxenon100} data put stringent constraint on the allowed range for dark matter mass. But the $N$ can still play the role of dark matter with appropriate masses.
\\

\begin{figure}
\includegraphics[width=6cm]{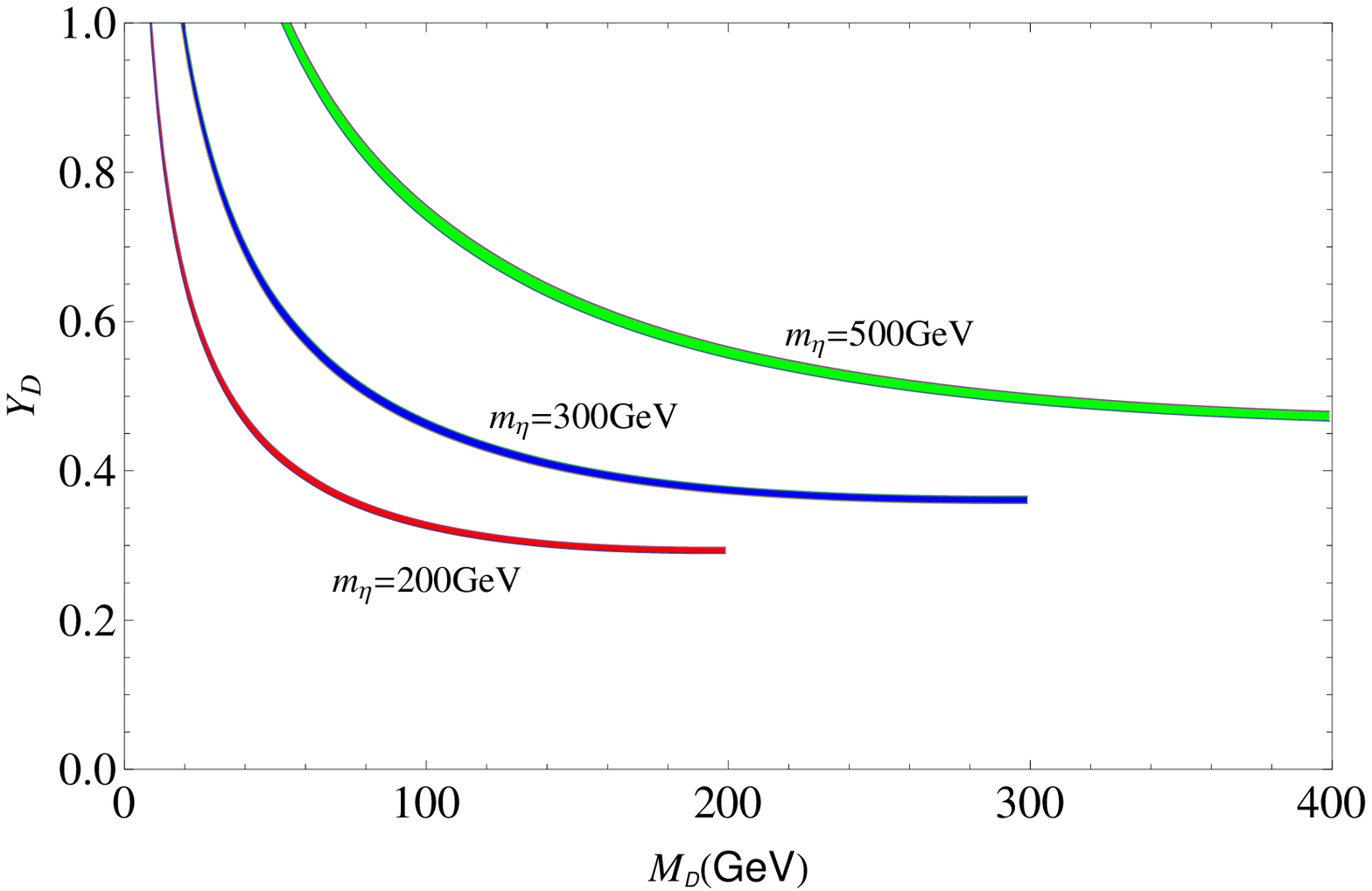}\\
\includegraphics[width=6cm]{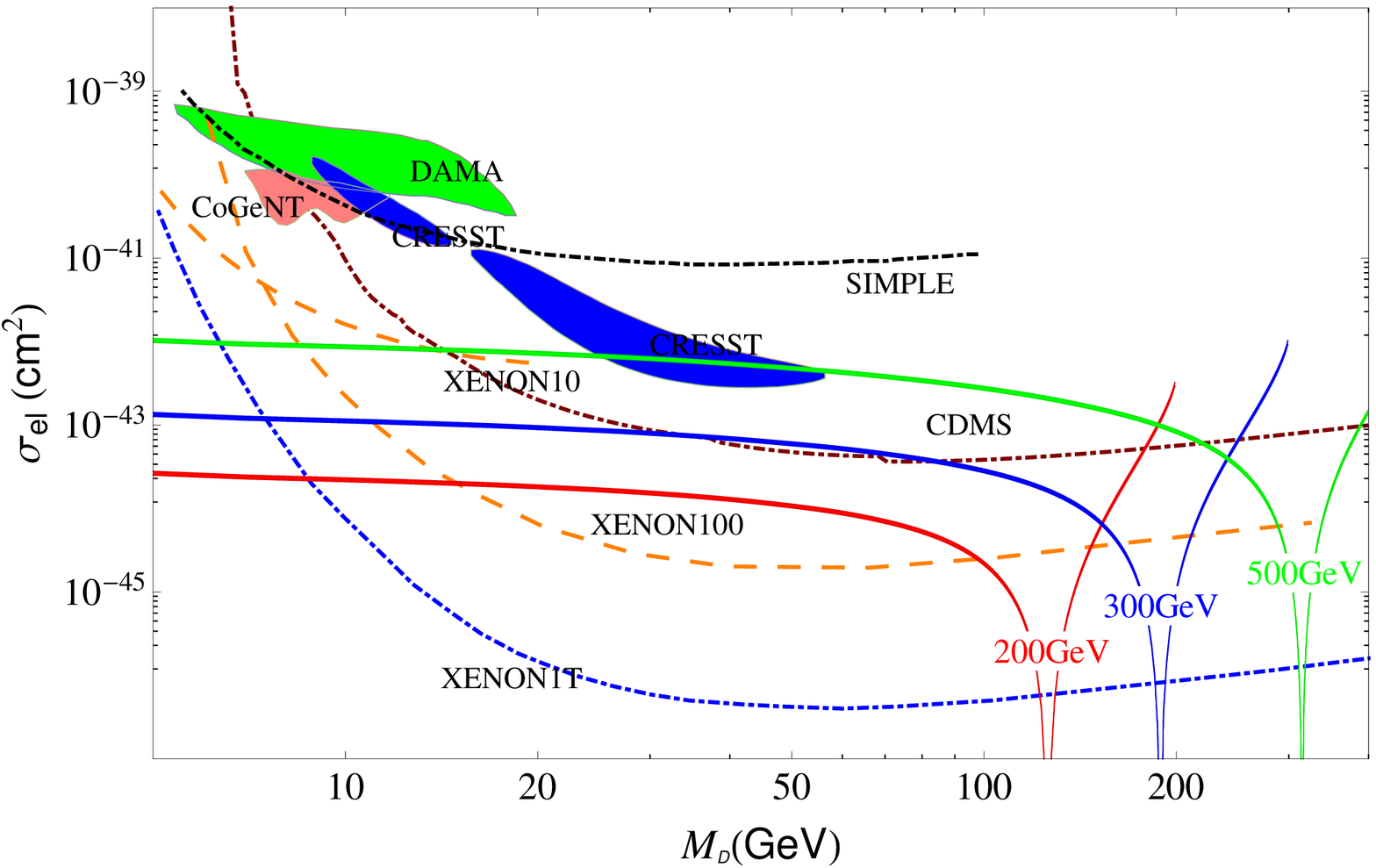}
\caption{Constraints on the coupling and dark matter mass from dark matter relic density and direct detection for $N$ as the dark matter. The different curves are for $\eta$ mass to be 200 GeV, 300 GeV and 500 GeV, respectively.}\label{N-dark-matter}
\end{figure}

\section{ Conclusions}

We have proposed two models, the $U(1)_D$ and $U(1)_S$ models, in which neutrino masses are generated through inverse seesaw mechanism at two loop level.
In these models, a global $U(1)$ is unbroken leading to a stable beyond SM new particle in each model. These stable new
particles are natural candidates for dark matter. We find that these models can satisfy current experimental constraints from neutrino masses, mixing, dark matter
relic density and direct detections.

Because in these models the neutrino masses are generated at two loop level and also inverse seesaw type,
the seesaw scale can be as low as a few hundred GeV. This can
lead to observable signatures. Before closing, we would like to make a few comments about some phenomenological implications of the models.

One of them is related to Higgs properties.
Although the Higgs couplings to SM particles are not modified at tree level, there are noticeable corrections at one loop level
in the above two models. An important example is the modification to $h \to \gamma\gamma$. Because the existence of the two terms
$(\Delta^\dagger \Delta H^\dagger H)_\alpha$ in both models and ($\eta^\dagger_i \eta_i H^\dagger_j H_j\;,\eta^\dagger_i \eta_j H^\dagger_j H_i)$
terms in the $U(1)_S$ model, terms like $\Delta^{++}\Delta^{--} h$ and $\Delta^{+}\Delta^{-} h$ ($\eta^+\eta^- h$) will be generated after $H$
develops vev, the $h\to \gamma\gamma$ can be modified.  At present the experimental value\cite{LHC-higgs} for this channel is $1.9\pm 0.5$ (ATLAS)
($1.56\pm 0.43$(CMS)) times that predicted by the SM. The central value is higher than the SM prediction.
With large enough $\lambda_{H\Delta }^\alpha$, one may bring the value to close to the data.

Another is related to probing the new degrees of freedom in the models at the LHC.
In both models there are new charged particles, the $\Delta$, $\eta$ and $D$ fields. The particles in this multiplet can be pair produced via
electromagnetic and weak interactions. However, in both models there are unbroken $U(1)$ symmetries,
the new particles cannot decay into pure SM final state making detection difficult. A possible signature is that the charged new particle decays
into an SM particle and a dark matter. The SM particle is detected, but the dark matter carries away large transverse missing momentum and energy.
For example for the $U(1)_D$ model, with $S$ been the dark matter, $D^\pm$ can decay into a charged lepton $l^\pm$ and the dark matter $S$.
In the $U(1)_S$ model with $N$ being the dark matter, $\eta^\pm$ can decay into a charged lepton and the dark matter.

Finally, there are potentially large FCNC effects in leptonic sector in these models. This is because that the Yukawa couplings $Y_D$ in both models can be of order $O(0.1)$,
at loop level exchange $D$ and $S$, and, $N$ and $S$ in the $U(1)_D$ and $U(1)_S$ models, respectively, can generate flavor changing radiative decay of
charged lepton $l\to l' \gamma$ with branching ratios close to the current experimental bound\cite{he-ren}. Also possible large $\mu \to e$ conversion.
Near future improved experiments can test these models\cite{he-ren}. Detailed analysis will be presented else where.

\acknowledgments \vspace*{-1ex}
This work was supported in part by NSC of ROC, and NNSF(grant No:11175115) and Shanghai science and technology commission (grant No: 11DZ2260700) of PRC. XGH would like to thank the Center for Theoretical Underground Physics and Related Areas (CETUP* 2012) in South Dakota for its hospitality and for partial support during the completion of this work.

\end{document}